\documentclass{ws-mplaMY}

\usepackage{graphicx}
\usepackage{hyperref}

\begin{document}

\markboth{A. A. Grib, Yu. V. Pavlov}
{On the limiting energy of the collision of elementary particles}

\title{\large On the limiting energy of the collision of elementary particles\\[4pt]
close to horizon of the rotating black hole}

\author{A. A. Grib}

\address{A.~Friedmann Laboratory for Theoretical Physics, Saint Petersburg, Russia\\
Theoretical Physics and Astronomy Department of the Herzen  University,\\
Moika 48, Saint Petersburg 191186, Russia\\
andrei\_grib@mail.ru}

\author{Yu. V. Pavlov}

\address{Institute of Problems in Mechanical Engineering of
Russian Academy of Sciences,\\
Bol'shoy pr. 61, St. Petersburg 199178, Russia\\
N.I.\,Lobachevsky Institute of Mathematics and Mechanics,
Kazan Federal University,\\
18 Kremlyovskaya St., Kazan 420008, Russia\\
yuri.pavlov@mail.ru}

\maketitle

\pub{Received (Day Month Year)}{Revised (Day Month Year)}

\begin{abstract}
    Arguments are given for the conclusion that the energy of collision of
two ultrarelativistic elementary particles due to gravitational radiation
cannot exceed the Planck value.
    Comparison of the gravitational and electromagnetic radiation for
charged particles close to the horizon of Kerr black hole is made.
    If for trans-Planckian energy the black hole can arise it is shown that
the energy growth used in scattering for interaction of particles is still
limited.

\end{abstract}

\ccode{PACS Nos.: 04.70.Bw, 97.60.Lf, 95.30.Sf, 11.80.-m}


\section{Introduction}
\label{secIntr}

    Particle collision in the vicinity of the horizon of the Kerr black hole
was considered in Refs.~\refcite{PiranSK75} and~\refcite{BSW}.
     In Ref.~\refcite{BSW}, a new resonance in two particle collision for
the case of extremal Kerr black hole was found (BSW effect).
     The authors claimed that the energy in the center mass reference frame
can be ``arbitrarily large'' growing up to infinity at the horizon.\cite{BSW}
    In our papers,\cite{GribPavlov2010,GribPavlov2011} the same effect was
found for nonextremal Kerr black holes in case of multiple scattering.
    In Refs.~\refcite{GribPavlov2010}--\refcite{BertiCardosoGPS09},
it was shown that not only the coordinate time but the proper time also needed
to get the infinite energy must be infinity.
    Later it was shown that such resonance occurs for more general case of
the charged and dirty rotating black holes.\cite{prd,Zaslavskii10b}
    The problem of particle collisions in the vicinity of black holes for
different models of black holes was also discussed in
Refs.~\refcite{Pourhassan12}--\refcite{Pourhassan19}.

    In Ref.~\refcite{GribPavlov2013}, another mechanism of getting the resonance
due to negative but large absolute value of the orbital momentum was
proposed for any point of the ergosphere.
    In Ref.~\refcite{GribPavlov2015} the resonance was obtained due to collision
close to the event horizon of the particle falling inside the black hole
with the particle moving from the black hole on the white hole geodesic.
    The need to account for gravitational radiation of a particle moving
around a rotating black hole in the calculation of collision energy was
indicated in the Ref.~\refcite{BertiCardosoGPS09}.
    In the case taking into account the gravitational radiation
of one object the authors of Ref.~\refcite{HaradaKimura11b} find that the
center-of-mass energy can be far beyond the Planck energy for dark matter
particles colliding near black holes.
     In Ref.~\refcite{TaZaslav13} and~\refcite{TaZaslav14}, it was shown that
under rather general and weak assumptions, the BSW effect survives even if
a force (modeling the effect of radiation of one particle etc.) acts on
the colliding particles.
    In Ref.~\refcite{Misyura}, gravitational radiation in collision of
superheavy particles is estimated in some special cases using
the Weinberg formalism.\cite{Weinberg}

    In this paper we give the answer to the question: ``Can the resonance
energy be made arbitrary high?''
    The answer occurs to be negative if one takes into account the role
of the gravitational radiation of the colliding particles.
    This radiation being negligible at low energy has the same order
as the resonance energy at the Planck energy and plays the role
of the ``bremsstrahlung'' radiation in electromagnetic case.
    Electromagnetic radiation of colliding charged particles is shown
to be much smaller than the gravitational radiation at high energies.

\section{The Energy of Collision and the Relative Velocity}
\label{secNRS}

    Let us find the energy $E_{\rm c.m.}$ in the center of mass system
of two colliding particles with rest masses~$m_1$ and~$m_2$
in arbitrary gravitational field.
    It can be obtained from
    \begin{equation} \label{SCM}
\left( E_{\rm c.m.}, 0\,,0\,,0\, \right) = m_1 c^2 u^i_{(1)} + m_2 c^2 u^i_{(2)},
\end{equation}
    where $c$ --- light velocity, $u^i=dx^i/ds$ --- 4-velocity.
    Taking the squared~(\ref{SCM}) and due to $u^i u_i=1$ one obtains
    \begin{equation} \label{SCM2af}
E_{\rm c.m.}^{\,2} = m_1^2 c^4 + m_2^2 c^4 + 2 m_1 m_2 c^4 u_{(1)}^i u_{(2) i}.
\end{equation}

    Let us find the  expression of the energy in the center of mass frame
through the relative velocity~$ v_{\rm rel}$ of particles at the moment
of collision.\cite{BanadosHassanainSilkWest10}
    In the inertial reference frame with the first particle at rest
at the moment of collision the four-velocity components are
    \begin{equation} \label{Relsk01}
u_{(1)}^i = (1,0,0,0), \ \
u_{(2)}^i = \!\left( \frac{c }{\sqrt{ c^2 - v_{\rm rel}^2}},\,
\frac{ \mathbf{v}_{\rm rel}}{\sqrt{ c^2 - v_{\rm rel}^2}} \right)\!.
\end{equation}
    So
    \begin{equation} \label{Relsk02}
u_{(1)}^i u_{(2) i} = \frac{1}{\sqrt{1- v_{\rm rel}^2/c^2}}, \ \
\frac{v_{\rm rel}}{c} =
\sqrt{1- \frac{\mathstrut 1}{\left( u_{(1)}^i u_{(2) i} \right)^2}}.
\end{equation}
    These expressions evidently don't depend on the coordinate system.

    From~(\ref{SCM2af}) and~(\ref{Relsk02}) one obtains
    \begin{equation} \label{Relsk03}
E_{\rm c.m.}^{\,2} = m_1^2 c^4 + m_2^2 c^4 +
\frac{2 m_1 m_2 c^4 }{\sqrt{1- v_{\rm rel}^2/c^2}}.
\end{equation}
    and the nonlimited growth of the collision energy in the center of mass
frame occurs due to growth of the relative velocity to the velocity of
light.

\section{Gravitational Radiation in Particle Scattering}
\label{secGI}

    At the moment of scattering of ultrarelativistic particles they move with
acceleration in variable gravitational field of each other.
    At this moment the gravitational radiation appears.
    In case of scattering of pointless classical massive particles S.\,Weinberg
formula (10.4.23) from Ref.~\refcite{Weinberg} gives the following expression
for the total energy per unit frequency interval emitted in gravitational
radiation in collision
    \begin{equation}
\frac{d E}{d \omega} = \frac{G}{2 \pi c} \sum_{N,M} \eta_N \eta_M m_N m_M
\frac{1+ \beta_{NM}^{\,2}}{\beta_{NM} (1 - \beta_{NM}^{\,2})^{1/2}}
\ln \left( \frac{1 + \beta_{NM}}{1 - \beta_{NM}} \right),
 \label{GIW}
\end{equation}
    where $G$ is the gravitational constant, $ N, M $ correspond to numeration
of particles in initial and final states and the factors $\eta_N$ are defined as
$$
\eta_N = \left\{
\begin{array}{l}
+1, \ \ N \ {\rm in \ final \ state} , \\
-1, \ \ N \ {\rm in \ initial \ state}  .
\end{array}
\right.
$$
    $\beta_{NM} = |\mathbf{v}_{NM}| / c$, \
$\mathbf{v}_{NM}$ is the relative velocity of particles with indexes~$N$ and~$M$.

    We use the approximation of the  Minkowski space-time because any Riemann
space on small distances can be considered to be flat and the scale of interaction
for collision of particles with high energies is small while the curvature of space
near the horizon of physical black holes (especially supermassive) is also small.
No special analysis of the structure of the horizon is needed.

    In Weinberg formula the effective potential of interaction is not present
but surely any change of the particle momentum is due to the interaction.
    For high energies close to the Planckian value this potential is unknown
which makes Weinberg formula especially useful.

    Evaluate the full energy in collision of two particles 
with ultrarelativistic energies
$$
E_{\rm c.m.} \gg m_1 c^2, m_2c^2 .
$$
    Then $\beta_{12} \sim 1$ and from~(\ref{Relsk03}) one obtains
    \begin{equation} \label{o1}
\sqrt{ 1- \beta_{12}^{\,2}} \approx \frac{2 m_1 m_2 c^4}{E_{\rm c.m.}^{\,2} },
\end{equation}
    \begin{equation} \label{o2}
1- \beta_{12} \approx 2 \left( \frac{m_1 m_2 c^4}{E_{\rm c.m.}^{\,2} } \right)^2.
\end{equation}
    Consider particles to be classical pointlike particles.
    In the sum~(\ref{GIW}) take $N,M = 1,2$ and $\beta_{12}$, $\beta_{21}$.
    For the energy of gravitational radiation in the unit frequency interval
in such collision one obtains
    \begin{equation} \label{GIWo}
\frac{d E}{d \omega} = \frac{2 G}{\pi c^5} E_{\rm c.m.}^{\,2}
\ln \left( \frac{E_{\rm c.m.}^{\,2} }{m_1 m_2 c^4} \right).
\end{equation}

    If one calculates the full radiated energy taking the integral of~(\ref{GIW})
one obtains as it is mentioned in Ref.~\refcite{Weinberg} the result divergent
as $\int^\omega d \omega$.
    This occurs due to the fact that formula~(\ref{GIW}) is obtained
(see Ref.~\refcite{Weinberg}) in approximation of immediate collision.
    In reality scatterings occur at some finite time interval  $\Delta t$
and so the integral in the frequency is cut at some value of $\omega$
of the order of $ 1/\Delta t$.
    If the collision occurs at the energy much less than
the Planck energy $ E_{\rm Pl} = \sqrt{\hbar c^5/G} = 1.22\cdot 10^{19}$\,GeV,
one can take as such frequency
    \begin{equation} \label{o3}
\omega \approx \frac{E_{\rm c.m.}}{\hbar}.
\end{equation}
    So one takes as the value of the frequency the inverse time for needed for
the light to pass the distance equal to the Compton length of the particle
with the energy $E_{\rm c.m.}$, i.e. $\lambda_C = c {\hbar} /E_{\rm c.m.}$.
    The corresponding time characterizes the duration of the intensive
interaction of the particles in the ultrarelativistic collision with
the impact parameter $\lambda_C $.
    To get the full radiation energy multiply~(\ref{GIWo}) on
the frequency~(\ref{o3}) and obtain
    \begin{equation}
E = \frac{2 G}{\pi c^5 \hbar} E_{\rm c.m.}^{\,3}
\ln \left( \frac{E_{\rm c.m.}^{\,2} }{m_1 m_2 c^4} \right)
=
\frac{4}{\pi} \frac{E_{\rm c.m.}^{\,3}}{E_{\rm Pl}^{\,2}}
\ln \left( \frac{E_{\rm c.m.}}{ E_{\rm Pl}}
\frac{ M_{\rm Pl}}{\sqrt{m_1 m_2}} \right),
 \label{GIWoI}
\end{equation}
    where $ M_{\rm Pl} = \sqrt{\hbar c/G} =2.18\cdot 10^{-8}$\,kg is Planck mass.

    If the value of $E_{\rm c.m.} $ is given the value of the logarithm in
the right hand side of~(\ref{GIWoI}) is not large even for such light particles
as electrons $\ln (M_{\rm Pl}/ m) < 52 $.
    So the role of the gravitational radiation in collisions of elementary particles
with the energy in the center of mass frame less than Planck energy is negligible
    \begin{equation} \label{GIWoIo}
E_{\rm c.m.} \ll E_{\rm Pl} \ \ \Rightarrow \\
\frac{E}{E_{\rm c.m.}} \ll 1 .
\end{equation}
    If the energy in the center of mass frame is large then due to
formula~(\ref{GIWoI}) the loss of energy thanks to gravitational radiation grows
as the cube of the energy and becomes comparable to the energy of particles
for $E_{\rm c.m.} \sim E_{\rm Pl}$.

    Note that the value of the gravitational radiation equal in order
with~(\ref{GIWoI}) can be obtained also from the well known
Einstein formula for the quadrupole radiation of gravitational waves
(see~(110.16) in Ref.~\refcite{LL_II})
    \begin{equation} \label{GIzLL}
- \frac{d E}{d t}  = \frac{G}{45c^5} \stackrel{...}{D}_{\alpha \beta}^{\,2}.
\end{equation}
    This formula can be used also for massless particles (photons).
    Take as the effective time $\Delta t = \hbar / E_{\rm c.m.} $
and the effective distance $\Delta r = \hbar c / E_{\rm c.m.} $.
    Such choice corresponds to the collision of particles taking into account
the De Broglie wave length for such energies.
    Putting in formula~(\ref{GIzLL})   
$\stackrel{...}{D}_{\alpha \beta} \to
(E_{\rm c.m.}/c^2) (\Delta r)^2/ (\Delta t)^3 = E_{\rm c.m.}^{\,2}/\hbar$,
one obtains for the energy of radiation of gravitational waves
    \begin{equation} \label{GIner}
E = \Delta t \left| \frac{dE}{d t} \right| =
\frac{1}{45} \frac{E_{\rm c.m.}^{\,3}}{E_{\rm Pl}^{\,2}}.
\end{equation}
    So the conclusion about the growth proportional to  $ E_{\rm c.m.}^{\,3} $
of the radiation energy if value $ E_{\rm c.m.} $ close to Planck energy
is valid in this approach also.

    Note that the situation for the electromagnetic radiation is different.
    To evaluate the effect use the formula for the intensity of
the electromagnetic radiation with terms of the second order in order
to take into account not only the dipole but also the quadrupole terms
to compare it with gravitational radiation(\ref{GIzLL}).
    Using formula~(71.5) from Ref.~\refcite{LL_II} one obtains
    \begin{equation} \label{GIzLLe}
- \frac{d E_{\rm em}}{d t}  = \frac{2}{3 c^3} \stackrel{..}{\mathbf{d}}^{2} +
\frac{1}{180 c^5} \stackrel{...}{\cal{D}}_{\alpha \beta}^{\,2} +
\frac{2}{3 c^3} \stackrel{..}{\bf \cal{M}}^{2},
\end{equation}
    where $ \mathbf{d} $ is electric dipole moment,
$ {\cal{D}}_{\alpha \beta} $ is electric quadrupole moment,
$ {\bf \cal{M}} = \sum q  [\mathbf{r} , \mathbf{v}] / (2 c) $
is the magnetic moment of the radiating system, $q$ is the particle charge.
    Calculating the electromagnetic radiation of the system of two colliding
ultrarelativistic particles with the energy $ E_{\rm c.m.} $ again take
time and distance $\Delta t = \hbar / E_{\rm c.m.} $,
$\Delta r = \hbar c / E_{\rm c.m.} $.
    Take the electric dipole moment as $ \mathbf{d} \sim e \Delta r$,
the electric quadrupole moment as
$ {\cal{D}}_{\alpha \beta} \sim e (\Delta r)^2 $,
the magnetic moment as $ {\cal{M}} \sim e  \Delta r /2 $,
where $e$ is the elementary charge.
    Then from~(\ref{GIzLLe}) one obtains
    \begin{equation}
- \frac{d E_{\rm em}}{d t}  = \frac{2}{3 c} \left( \frac{e}{\Delta t} \right)^2
+ \frac{1}{180 c} \left( \frac{e}{\Delta t} \right)^2 +
\frac{1}{6 c} \left( \frac{e}{\Delta t} \right)^2
= \frac{151}{180 c } \left( \frac{e E_{\rm c.m.}}{\hbar } \right)^2.
\label{GIzLLeO}
\end{equation}
    Multiplying both sides on the effective time one obtains full energy of
the electromagnetic radiation in ultrarelativistic collision of particles
with size of the De Broglie wave length at such energies
    \begin{equation} \label{GInEM}
E_{\rm em} = \Delta t \left| \frac{d E_{\rm em}}{d t} \right| =
\frac{151}{180} \frac{e^2}{\hbar c} E_{\rm c.m.} \approx
\frac{1}{137} \cdot \frac{151}{180}  E_{\rm c.m.}.
\end{equation}
    So the relation of the energy of electromagnetic radiation to the energy
of collision in the center of mass frame has the approximate value equal
to the value of the fine structure constant $e^2 / (\hbar c) \approx 1/137$.

    For small energies the electromagnetic radiation is larger than the
gravitational in many times.
    However due to the growth of the gravitational interaction in relativistic
region proportional to the energies of colliding particles the gravitational
radiation grows in energy greater than the electromagnetic one and as
it is seen from our estimates becomes dominant close to the Planck energy.
    Comparing formulas~(\ref{GIWoI}) and (\ref{GInEM}) one obtains that
the dominance of the gravitational radiation over the electromagnetic one
begins from values $E_{\rm c.m.} \approx \sqrt{e^2/ (\hbar c)} \, E_{\rm Pl} $.

    Evaluate the power of radiation of gravitational waves taking the time
$\Delta t = 2 \pi \hbar / E_{\rm c.m.} $ in~(\ref{GIWoI}).
    Then one obtains
    \begin{equation} \label{GIWoN}
\frac{\Delta E}{\Delta t} = P_{\rm Pl} \left( \frac{E_{\rm c.m.}}{E_{\rm Pl}} \right)^4
\frac{8 }{\pi^2} \ln \left( \frac{E_{\rm c.m.}}{ E_{\rm Pl}}
\frac{ M_{\rm Pl}}{\sqrt{m_1 m_2}} \right),
\end{equation}
    where $P_{\rm Pl} = c^5/(4G) \approx  9.07 \cdot 10^{51}$\,W is the Planck power.
    So for $ E_{\rm c.m.} \ll E_{\rm Pl} $,
the radiating power is much smaller than the Planck one but when the energy
of collision grows close to the Planck value the power of radiation also grows
to the Planck power $P_{\rm Pl} $.

    In case when the energy of elementary particles in the center of mass frame
is larger than the Planck value formula~(\ref{GIWoI}) shows that practically
all energy will be radiated in the form of gravitational waves.
    If particles come close one to another on the distance of the Compton
order~$\lambda_C$ the remaining energies not higher than the Planckian one.

    Note that for trans-Planckian energies the Compton wave length becomes
smaller than the Planckian length
$l_{\rm Pl} = \sqrt{ \hbar G / c^3} \approx 1.6 \cdot 10^{-35}$\,m
and the corresponding frequency~(\ref{o3}) will be larger than
the Planckian frecuency
$ \omega_{\rm Pl} = E_{\rm Pl} /\hbar = \sqrt{c^5 / \hbar G} = 1.85 \cdot 10^{43}$\,s$^{-1}$.
    The power of radiation in the form of gravitational waves in
scattering with such trans-Planckian parameters becomes due to~(\ref{GIWoN})
larger than the Planckian power.

    However one can put the hypothesis that some ``black hole'' with
radius of events horizon of order $r_H = G E_{\rm c.m.} /c^4 $  can be formed
(see Refs.~\refcite{Hooft87})
for trans-Planckian energies.
    Then a new parameter is arising leading to the frequency
    \begin{equation} \label{o4p}
\omega_H = \frac{c}{r_H} = \frac{c^5}{E_{\rm c.m.} G},
\end{equation}
    and to the new corresponding impact parameter $r_H $ of the particles collision
one obtains multiplying~(\ref{GIWo}) on
the frequency~(\ref{o4p}) another formula for the full energy of
gravitational radiation
    \begin{equation} \label{pnn}
E \approx \frac{4}{\pi} E_{\rm c.m.} \ln \left( \frac{E_{\rm c.m.}}{ E_{\rm Pl}}
\frac{ M_{\rm Pl}}{\sqrt{m_1 m_2}} \right).
\end{equation}
    Then gravitational radiation occurs to be smaller than in case~(\ref{GIWoI}).
    Obviously, the using of the formula describing the collision of point
particles can give only an estimate of the radiation energy in order of
magnitude in the region of such parameter.
    Thus, the formula~(\ref{pnn}) means that energy of gravitation radiation
is equal in order to the collision energy $ E_{\rm c.m.} $.

    If the impact parameter~$b$ is very small $b < r_H$ then two particles
disappear inside of the horizon of the effective nonextremal Kerr's
black hole.
    It's angular momentum~$J$ for the impact parameter is $ E_{\rm c.m.} b /c $
and it is less than maximal angular momentum $J_{\rm max}$
for Kerr's black hole with mass $ E_{\rm c.m.} / c^2 $:
    \begin{equation} \label{ehK}
b < r_H \ \ \ \Rightarrow \ \ \ J= \frac{E_{\rm c.m.} b}{c} <
J_{\rm max} = \frac{G E_{\rm c.m.}^2}{c^5}.
\end{equation}
    So there is no scattering at all and Weinberg's formula is not valued.

    That is why if scattering is present the impact factor must be
larger than~$r_H$.
    For this case one can evaluate the energy of each particle after scattering
in the presence of a mini black hole.
    Their energy energy can be evaluated as the result of interaction of the particle
with the rotating Kerr's black hole near the event horizon $ r_H$.
    Any test particle near the horizon has the angular velocity equal to
the angular velocity of the rotating black hole
    \begin{equation} \label{OmBh}
\Omega_{Bh} = \frac{J_{Bh} c^2}{2 G M^2 r_H},
\end{equation}
    where $J_{Bh}$ is the angular momentum of the black hole.
    The force acting from the point of the distant observer on the particle
with 3-momentum $\mathbf{p}$ is equal to
    \begin{equation} \label{OmF}
F= \left| \frac{d \mathbf{p}}{d t} \right| = p \Omega_{Bh},
\end{equation}
    where $p = |\mathbf{p}|$.
    In the center of mass frame $p \approx E_{\rm c.m.}/(2 c) $.
    Note that the maximal value of this force obtained by putting
into(\ref{OmF}) $J=J_{\rm max}$ (see~(\ref{ehK})) is
    \begin{equation} \label{OmFmax}
F_{\rm max} =  \frac{c^4}{4 G} \approx 3.026 \cdot 10^{43}\,{\rm N} . 
\end{equation}
    It is equal to well known value of the maximal
force.\cite{Gibbons02,Schiller05}

    The power of the interaction relative to distant observer at rest is
    \begin{equation} \label{OmP}
P = \mathbf{F} \mathbf{v} \le F c = p \Omega_{Bh} c,
\end{equation}
    where $\mathbf{v}$ is 3-velocity relative to distant observer.
    Due to~(\ref{OmBh}) one obtains
    \begin{equation} \label{OmPm}
P \le \frac{p J c^3}{2 G M^2 r_H}.
\end{equation}
    The limitation for the interaction intensity of colliding particles with
the impact parameter $r_H$ one obtains by putting
$M = E_{\rm c.m.}/c^2$, $p = E_{\rm c.m.}/(2 c) $ and $J=J_{\rm max}$:
    \begin{equation} \label{OmPmm}
P \le \frac{c^5}{4 G } = P_{\rm Pl}.
\end{equation}

    The power of the interaction force for the energies of collision in
the center of mass frame smaller than the Planckian one with interaction time
$ 2 \pi \hbar / E_{\rm c.m.} $ (see~(\ref{o3})) has the order
    \begin{equation} \label{WGPcom}
P = \frac{E_{\rm c.m.}^2}{h} = \left(\frac{E_{\rm c.m.}}{ E_{\rm Pl} }\right)^2
\frac{2}{\pi} P_{\rm Pl}.
\end{equation}
    So in the scattering of particles with any energy even in case of
the formation of the mini black hole the intensity of interaction cannot be
larger than the Planckian power~$P_{\rm Pl}$.

    Surely one can reasonably note that consideration of Planck time and
Plank power lead to quantization of gravity and then there is no sense in
such notions as the critical trajectory etc., leading to
``super Planck high energy resonance''.
    This is again the argument for existence of the limit of the energy
of the resonance discussed in this paper.

\section{Conclusion}
\label{Conclus}

    Some people before the experimental registration of gravitational
waves (see Ref.~\refcite{GravWave16}) asked:
``Can the observable Universe exist without gravitational waves?
What is the necessity of their existence?''
    The important conclusion of our paper is that gravitational radiation
can help be free of the inconsistency arising from nonlimited growth of
the energy in collisions close to the horizon of black holes
due to gravitational radiation playing the role of
the ``bremsstrahlung''.\cite{Ciafaloni15,Gruzinov16}
    It is interesting to note that Einstein's formula without any use of
impact parameters of scattering gives the same result as Weinberg formula.
    For small impact parameter in spite the result for the gravitational
wave is different the growth of the energy
of particle interactions is still occurs to be limited.

\vspace*{-1mm}
\section*{Acknowledgments}
This research is supported by the Russian Foundation for Basic research
(Grant No. 18-02-00461 a).
The work of Yu.V.P. was supported by the Russian Government Program of
Competitive Growth of Kazan Federal University.


\end{document}